\documentclass[ aip, jmp, amsmath,amssymb, reprint, ]{revtex4-1}

\usepackage{amssymb}
\usepackage{amsfonts}
\usepackage{amsmath}
\usepackage{amsthm}
\usepackage{epsfig}
\usepackage{color} 
\usepackage{subfigure}

\graphicspath{{./}{./Figures/}}

\begin{document}

\title{Nonlinearity of local dynamics promotes multi-chimeras}

\author{Iryna Omelchenko}
\email{omelchenko@itp.tu-berlin.de}
\affiliation{Institut f{\"u}r Theoretische Physik, Technische Universit{\"a}t Berlin, Hardenbergstra\ss{}e 36, 10623 Berlin, Germany}
\author{Anna Zakharova}
\email{anna.zakharova@tu-berlin.de}
\affiliation{Institut f{\"u}r Theoretische Physik, Technische Universit{\"a}t Berlin, Hardenbergstra\ss{}e 36, 10623 Berlin, Germany}
\author{Philipp H{\"o}vel} 
\email{phoevel@physik.tu-berlin.de}
\affiliation{Institut f{\"u}r Theoretische Physik, Technische Universit{\"a}t Berlin, Hardenbergstra\ss{}e 36, 10623 Berlin, Germany}
\affiliation{Bernstein Center for Computational Neuroscience Berlin, Humboldt-Universit{\"a}t zu Berlin,
Philippstra{\ss}e 13, 10115 Berlin, Germany}
\author{Julien Siebert}
\email{j.siebert@mailbox.tu-berlin.de}
\affiliation{Institut f{\"u}r Theoretische Physik, Technische Universit{\"a}t Berlin, Hardenbergstra\ss{}e 36, 10623
Berlin, Germany}
\author{Eckehard Sch{\"o}ll}
\email{schoell@physik.tu-berlin.de}
\affiliation{Institut f{\"u}r Theoretische Physik, Technische Universit{\"a}t Berlin, Hardenbergstra\ss{}e 36, 10623 Berlin, Germany}

\date{\today}

\begin{abstract}
Chimera states are complex spatio-temporal patterns in which domains of synchronous and asynchronous dynamics coexist in
coupled systems of oscillators. We examine how the character of the individual elements influences chimera states by
studying networks of nonlocally coupled Van der Pol oscillators. Varying the bifurcation parameter of the Van der Pol
system, we can interpolate between regular sinusoidal and strongly nonlinear relaxation oscillations, and demonstrate
that more pronounced nonlinearity induces multi-chimera states with multiple incoherent domains. We show that the
stability regimes for multi-chimera states and the mean phase velocity profiles of the oscillators change significantly
as the nonlinearity becomes stronger. Furthermore, we reveal the influence of time delay on chimera patterns.

\end{abstract}

% Synchronization, nonlinear dynamics, 05.45.Xt
% Complex systems, 89.75.-k

\maketitle

%Lead paragraph
\begin{quotation}
The investigation of coupled oscillatory systems is an important research field bridging between nonlinear dynamics, network science, and
statistical physics, with a variety of applications in physics, biology, and technology~\cite{PIK01,BOC06a}. The analysis and numerical simulation of large networks with complex coupling schemes continues to open up new unexpected dynamical scenarios. Chimera states are an example for such intriguing phenomena; they exhibit a hybrid structure combining coexisting domains of both coherent (synchronized) and incoherent (desynchronized) dynamics, and were first reported for the well-known model of phase oscillators~\cite{KUR02a,ABR04}. 
In this paper, we investigate the influence of the local dynamics of the oscillators upon the resulting chimera
patterns. Using the
Van der Pol oscillator, which is a model allowing for a continuous transition between sinusoidal and strongly nonlinear relaxation oscillations by tuning a single parameter, we show that multi-chimera patterns with multiple incoherent domains are promoted by increasing the nonlinearity of the local oscillator dynamics.
\end{quotation}

\section{Introduction}
The last decade has seen an increasing interest in chimera states in dynamical
networks \cite{LAI09,MOT10,MAR10,OME10a,OME12a,MAR10b,WOL11a,BOU14,PAN15}. It was shown that they are not limited to
phase oscillators, but can be found in a large variety of different systems including time-discrete maps~\cite{OME11},
time-continuous chaotic models~\cite{OME12}, neural systems~\cite{OME13,HIZ13,VUE14a}, and Boolean networks
\cite{ROS14a}. 
Moreover, chimera states were found in systems with higher spatial dimensions~\cite{OME12a,SHI04,MAR10,PAN13,PAN15,PAN15a}.
Together with the initially reported chimera states, which consist of one coherent and one incoherent domain, new types
of these peculiar states having multiple incoherent regions~\cite{SET08,OME13,MAI14,XIE14,VUE14a}, as well as
amplitude-mediated~\cite{SET13,SET14}, and pure amplitude chimera and chimera death states~\cite{ZAK14} were discovered.

In many systems, the form of the coupling defines the possibility to obtain chimera states. The nonlocal coupling has
generally been assumed to be a necessary condition for chimera states to evolve in coupled systems. However, recent
studies have shown that even global all-to-all coupling~\cite{SET14,YEL14,SCH14g,BOE15}, as well as more complex
coupling topologies allow for the existence of chimera states~\cite{KO08,SHA10,LAI12,YAO13,ZHU14,OME15}. Furthermore,
time-varying network structures can give rise to alternating chimera states \cite{BUS15}.

The important question of the main features that give rise to chimera states in coupled systems has been widely discussed, but no conclusive answer has been given yet. 
In systems of phase oscillators, the value of the phase lag parameter $\alpha$, which occurs in the coupling function, is crucial. In nonlocally coupled systems, the range of the coupling and its strength play the key role. If the local dynamics of each unit is described by a two- or higher dimensional system, then the interaction scheme between the units plays an important role, i.e., which variable is coupled to which variable of the other nodes. Chimera states have also been shown to be robust against inhomogeneities of the local dynamics and coupling topology~\cite{LAI10,OME15}.

Possible applications of chimera states in natural and technological systems include the phenomenon of unihemispheric sleep~\cite{RAT00}, bump states in neural systems~\cite{LAI01,SAK06a}, power grids~\cite{FIL08}, or social systems~\cite{GON14}.  
Many works considering chimera states have mostly been based on numerical results. A deeper bifurcation analysis~\cite{OME13a} and even a possibility to control chimera states~\cite{SIE14c,BIC14} were obtained only recently.

The experimental verification of chimera states was first demonstrated in optical~\cite{HAG12} and chemical~\cite{TIN12,NKO13} systems. Further experiments involved  mechanical~\cite{MAR13}, electronic~\cite{LAR13,GAM14} and electrochemical~\cite{SCH14a, WIC13} oscillator systems, Boolean networks~\cite{ROS14a}, the optical comb generated by a passively mode-locked quantum dot laser~\cite{VIK14}, and superconducting quantum interference devices~\cite{LAZ15}. 

In previous investigations of chimera states, usually the character of the local node dynamics has been
considered as fixed. In the current study, we address the issue of the impact of the local dynamics. We analyze the
properties of chimera states, when the dynamics of individual
oscillators smoothly changes from sinusoidal to nonlinear relaxation oscillations. For this reason, we choose the Van
der Pol oscillator to describe the dynamics of each node.  The Van der Pol oscillator~\cite{POL26} has a long history of
being used in both the physical and biological sciences, as a generic model for electrical circuits~\cite{ZAK13} and
action potentials of neurons, respectively. 
%*******************************************************
\begin{figure*}[ht!]
\includegraphics[height=\linewidth, angle=270]{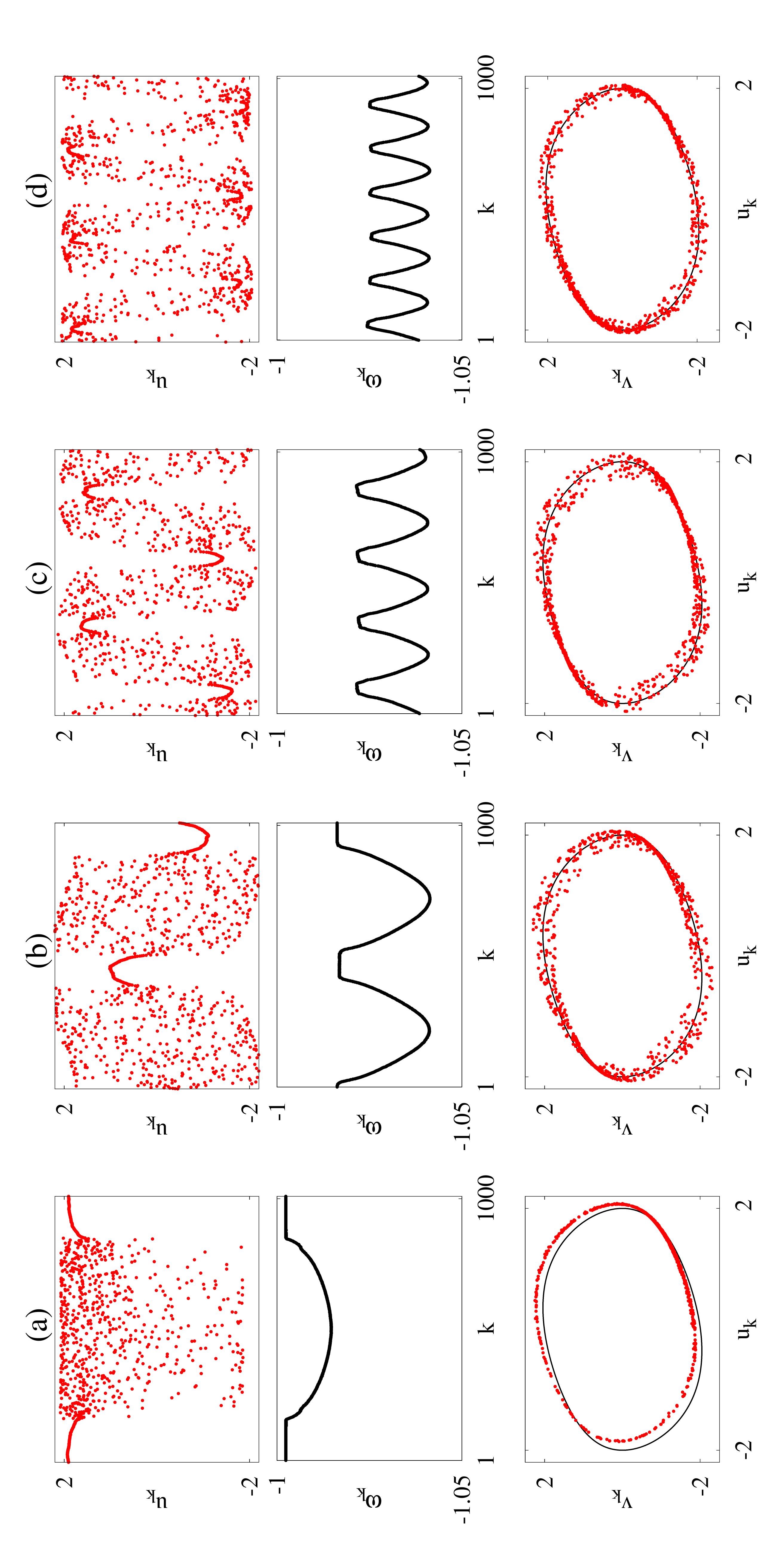}
\caption{Snapshots of the variables $u_k$ (upper panels), mean phase velocities $\omega_k$ (middle panels), and snapshots in the phase space $(u_k,v_k)$ (bottom panels, limit cycle of the uncoupled unit shown black).  (a)~$r=0.35$, $\sigma=0.05$, (b)~$r=0.2$, $\sigma=0.09$, (c)~$r=0.13$, $\sigma=0.09$, (d)~$r=0.1$, $\sigma=0.09$. Other parameters: $N=1000$, $b_1 = 1$, $b_2 =0.1$, and $\varepsilon =0.2$.}
\label{fig1}
\end{figure*}
%*******************************************************

\section{The model}
\label{sec:model}

In our study, we consider a system of nonlocally coupled Van der Pol oscillators with ring topology, where each element
of the system interacts with a fixed range of its neighbors in both directions:
\begin{equation}
\begin{aligned}
\ddot{u}_k = & \varepsilon (1-u_k^2)\dot{u}_k - u_k \\
& + \dfrac{\sigma}{2R} \sum\limits_{j=k-R}^{j=k+R} \left[ b_1(u_j-u_k) + b_2(\dot{u}_j-\dot{u}_k)\right],\\
\end{aligned}
\label{Eq1}
\end{equation}
with $k=1,...,N$ where  all indices are taken modulo $N$, $\varepsilon$ is the bifurcation parameter of the individual
oscillator, $\sigma$~denotes the strength of the coupling, $R$ is the number of coupled neighbors (in each direction), 
and $b_1,b_2$ are the interaction parameters.
For such a form of coupling it is convenient to consider the ratio $r=R/N$, which we denote as a coupling range.
The uncoupled Van der Pol oscillator has a stable trivial steady state $u=0$ for $\varepsilon<0$ and exhibits a supercritical Hopf bifurcation at $\varepsilon=0$. Here we consider $\varepsilon>0$.

Introducing a new variable $ v_k=\dot{u}_k$, Eq.~(\ref{Eq1}) can be rewritten in the form of a two-dimensional system: 
\begin{equation}
\begin{aligned}
\dot{u}_k = & v_k\\
\dot{v}_k = & \varepsilon (1-u_k^2)v_k - u_k \\
& +\dfrac{\sigma}{2R} \sum\limits_{j=k-R}^{j=k+R} \left[ b_1(u_j-u_k) + b_2(v_j-v_k)\right].\\
\end{aligned}
\label{Eq2}
\end{equation}

The form of the coupling in the system~Eq.~(\ref{Eq1}) or (\ref{Eq2}) is inspired from biological systems, describing interaction of the cells or pattern generation in locomotion~\cite{LOW03,LOW06}. A similar form of the coupling is also used in mechanics~\cite{STO00}. The cross-couplings between the $u$- and the $v$-variable play an important role, they were shown to be necessary for the existence of chimera and multi-chimera states in systems of nonlocally coupled FitzHugh-Nagumo oscillators~\cite{OME13}.

The dynamics of the system Eq.~(\ref{Eq2}) is determined by five parameters: $\varepsilon$ defines the dynamics of each
individual unit, and the parameters $\sigma$, $R$, $b_1$, and $b_2$ specifies the coupling. In order to find suitable
values for some of the system parameters in the regime where Eq.~(\ref{Eq2}) can describe chimera states, we will use
the experience from simpler systems of coupled Kuramoto phase oscillators.  For this reason we transform our system using the phase averaging technique on a rotating frame for slowly varying amplitude $r_k$ and phase
$\theta_k$: $u_k(t)=r_k(t) \sin (t+\theta_k(t))$ and $v_k(t)=r_k(t) \cos (t+\theta_k(t))$. As a result, we obtain the approximate system 
\begin{equation}
\label{Eq_phaseampl}
\begin{aligned}
 \dot{r}_k = & \dfrac{\varepsilon}{8}r_k \left[ \left( 4-\dfrac{2\sigma}{\varepsilon R} (2R+1)b_2 \right)-r_k^2 \right]
 \\
 & +\dfrac{\sigma}{4R} \sqrt{b_1^2 + b_2^2} \sum\limits_{j=k-R}^{k+R} r_j \cos(\theta_k - \theta_j + \alpha)\\
 \dot{\theta}_k = & \dfrac{\sigma}{4R} (2R+1)b_1 \\
 & -\dfrac{\sigma}{4R} \sqrt{b_1^2+b_2^2} \sum\limits_{j=k-R}^{k+R}\dfrac{r_j}{r_k} \sin(\theta_k - \theta_j + \alpha) 
\end{aligned}
\end{equation}
with $\alpha=\arctan (b_1/b_2)$, $b_2>0$, and  $k=1,...,N.$ 

The parameter $\alpha$ in the system~(\ref{Eq_phaseampl}) can be associated with the phase lag parameter in the systems
of coupled phase oscillators~\cite{ABR04}. This parameter is crucial for the appearance of chimera states in the phase
oscillator network. In~\cite{OME10a} it was shown that a value of the phase lag parameter close to but slightly
less than $\pi/2$ allows for the existence of chimera states. 

%*******************************************************
\begin{figure*}[ht!]
\includegraphics[height=0.8\linewidth, angle=270]{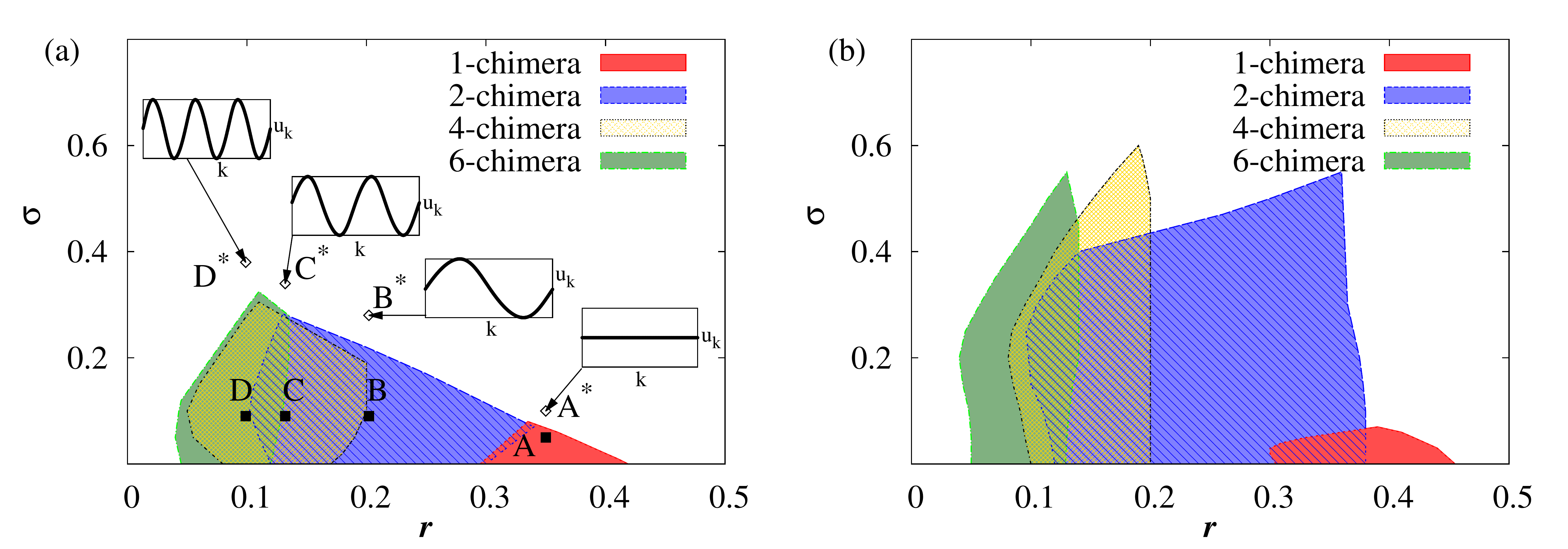}
\caption{Stability regimes for multiple chimera states.
% $N=1000$, $b_1 = 1$, $b_2 =0.1$. 
(a)~$\varepsilon =0.2$, black squares marked by A-D show parameter values corresponding
to panels (a)-(d) in Fig.~\ref{fig1}. The insets show snapshots of coherent spatial profiles for parameter values $A^*(r=0.35, \sigma=0.12; K=0)$, $B^*(r=0.2, \sigma=0.28; K=1)$, $C^*(r=0.13, \sigma=0.34; K=2)$, $D^*(r=0.1, \sigma=0.38; K=3)$; (b)~$\varepsilon =0.4$. Other parameters as in Fig.~\ref{fig1}.}
\label{fig2}
\end{figure*}
%*******************************************************

In the following, using the experience from the phase oscillator network, we fix the interaction parameters to be 
$b_1=1$ and $b_2=0.1$, such that $\alpha\approx 1.47$ is close to $\pi/2$. With this parameter choice,
we will focus further on the original system Eq.~(\ref{Eq2}), and vary the parameter~$\varepsilon$ that defines the type
of local dynamics of each element, as well as the coupling parameters $\sigma$ and $r$ describing the strength and the
range of the coupling, respectively. 

\section{The impact of local dynamics}
\label{sec:impact}

Varying the bifurcation parameter~$\varepsilon$ results in a change of the character of the local node dynamics.
If~$\varepsilon$ is small, the uncoupled individual elements of the system perform harmonic oscillations on a limit
cycle, which is approximately a circle. With increasing~$\varepsilon$, the individual limit cycle becomes distorted and
changes its form to relaxation oscillations.

Figure~\ref{fig1} demonstrates examples of chimera states for the system of $N=1000$ elements, $\varepsilon=0.2$, and decreasing coupling range. The upper panels depict snapshots of the variables $u_k$ for fixed time $T=50000$. As initial conditions we use randomly distributed phases on the circle $u^2+v^2=4$, i.e., around the limit cycle of the uncoupled system, which is approximately a circle of radius~$2$. 
One can clearly distinguish coherent and incoherent domains, a characteristic signature of chimera states. 
Elements that belong to the incoherent domain are scattered along the limit cycle, as shown with red points in the bottom panels of Fig.~\ref{fig1} where the black line denotes the limit cycle of the uncoupled unit with corresponding value $\varepsilon=0.2$. 
The individual nodes perform a nonuniform rotational motion, but neighboring oscillators are not phase-locked. To illustrate this, the middle panels of Fig.~\ref{fig1} show the mean phase velocities for each oscillator calculated as $\omega_k = 2 \pi M_k/\Delta T$, $k=1,...,N$, where $M_k$ is the number of complete rotations around the origin performed by the $k$-th node during the time interval $\Delta T$. Throughout the paper we use $\Delta T=50000$ for the calculation of the mean phase velocities~$\omega_k$ corresponding to several thousand periods. The values of $\omega_k$ lie on a continuous curve and the interval of constant $\omega_k$ corresponds to the coherent domain, where neighboring elements are phase-locked. This mean phase velocity profile is a clear indication of chimera states and similar to the case of coupled Kuramoto phase oscillators~\cite{KUR02a,ABR04}.

In addition to chimera states with one incoherent domain [Fig.~\ref{fig1}(a)], we observe chimera states with multiple
incoherent domains shown in Fig.~\ref{fig1}(b)-(d), i.e., multi-chimera states. In the following, we will use the
notation $n$-chimera for a chimera state with $n$ coherent and $n$ incoherent domains.  In analogy with networks of
phase oscillators, here we observe chimera states where the number of incoherent domains is even. The number of
incoherent domains increases with decreasing coupling range.  

Figure~\ref{fig2}(a) shows the stability regimes for chimera states with one and multiple incoherent domains in the
plane of coupling range~$r$ and coupling strength~$\sigma$ for $\varepsilon=0.2$.  Indeed, for large coupling range, we
observe the stability regime for chimera states with one incoherent domain, and regimes for chimera states with two,
four, and six incoherent domains follow subsequently with decreasing coupling range. The overlaps of these regimes are
characterized by multistability, when each of the chimera states can be obtained in the system depending on the choice
of the initial conditions. The regimes shown in the diagram are obtained by starting from the chimera states shown in
Fig.~\ref{fig1}, and using this pattern as initial condition for the neighboring parameter set, and so forth with a step size of $\Delta r=0.01$ and $\Delta\sigma=0.01$. Black
squares denoted by A-D show values of the parameters $(r,\sigma)$ that correspond to the examples presented in
Fig.~\ref{fig1}(a)-(d), respectively.

For larger coupling strength~$\sigma$, we observe coherent states in the system~(\ref{Eq2}). They are characterized by the wavenumber~$K$ defining the number of maxima (minima) in the spatial profile, and $K=0$ corresponds to complete in-phase synchronization. The wavenumber increases with decreasing coupling radius, and exemplary snapshots are shown in the insets of Fig.~\ref{fig2}(a). For large coupling strength~$\sigma$, the system is characterized by high multistability, and depending upon initial conditions one can obtain coherent solutions with different wavenumber. 
%The transition between the completely coherent solutions and chimera states occurs via coherence-incoherence bifurcation, where the coherent profile breaks down into several parts (depending on its wavenumber) with further appearance of the incoherent domains in between, as described in~\cite{OME11}.
%%%
In our system there exist two different types of chimeras, amplitude-mediated chimeras and pure phase chimeras, and these are generated by different bifurcation mechanisms. The amplitude-mediated 2-, 4-, 6-chimeras (Fig.~2(a), snapshots in Fig.~1(b),(c),(d), top panel) are generated from smooth, completely coherent spatial profiles of wavenumbers $K=1,2,3$, respectively, by a coherence-incoherence bifurcation with decreasing coupling strength as indicated by the insets in Fig.~2(a). 
%Here the integer $K=1,2,3,...$ labels the spatial period of the corresponding coherent profile, i.e. the spatial wavelength is $1/K$. 
At the onset of chimeras the smooth coherent profiles break up into spatially coherent domains corresponding to the upper and the lower parts of these profiles, and incoherent domains in between. Therefore these incoherent domains occur in pairs (2-, 4-, 6-chimeras). Such coherence-incoherence bifurcations have also been observed for other local dynamics, e.g., logistic maps and R{\"o}ssler systems \cite{OME11}, the cosine map \cite{HAG12}, and Stuart-Landau oscillators \cite{ZAK14}. In contrast, the pure phase chimeras (1-chimera in Fig.~2(a),(b))
%, and 4-,6-,8-chimeras in points F,G,H,I,J of Fig.3) 
arise from completely in-phase synchronized ($K=0$) profiles. %, and outside the chimera regime in the $(r,\sigma)$ control parameter space we have only found completely synchronized or completely chaotic solutions. 
The shift of the regimes for (multiple) $n$-chimeras to smaller coupling range $r$ with increasing $n$
is typical for various nonlocally coupled systems \cite{OME11,OME12,OME13,ZAK14,HAG12}.
% and has been corroborated analytically by a scaling analysis in the $N \to \infty$ limit \cite{OME12,HAG12}.
%%%
 
For small~$\varepsilon$, the limit cycle of each individual Van der Pol oscillator is close to a circle, corresponding to sinusoidal oscillations, and the similarities to the chimera states in a system of phase oscillators are clearly revealed. 
However, the hybrid solutions we observe in the system of coupled Van der Pol oscillators demonstrate chimera behavior both for phases and amplitudes. This can be seen in the bottom panels of Fig.~\ref{fig1}, where red dots denoting the snapshot  of all nodes deviate in their amplitudes slightly from the limit cycle of the uncoupled unit (black line).

%*******************************************************
\begin{figure*}[ht!]
\includegraphics[height=0.8\linewidth, angle=270]{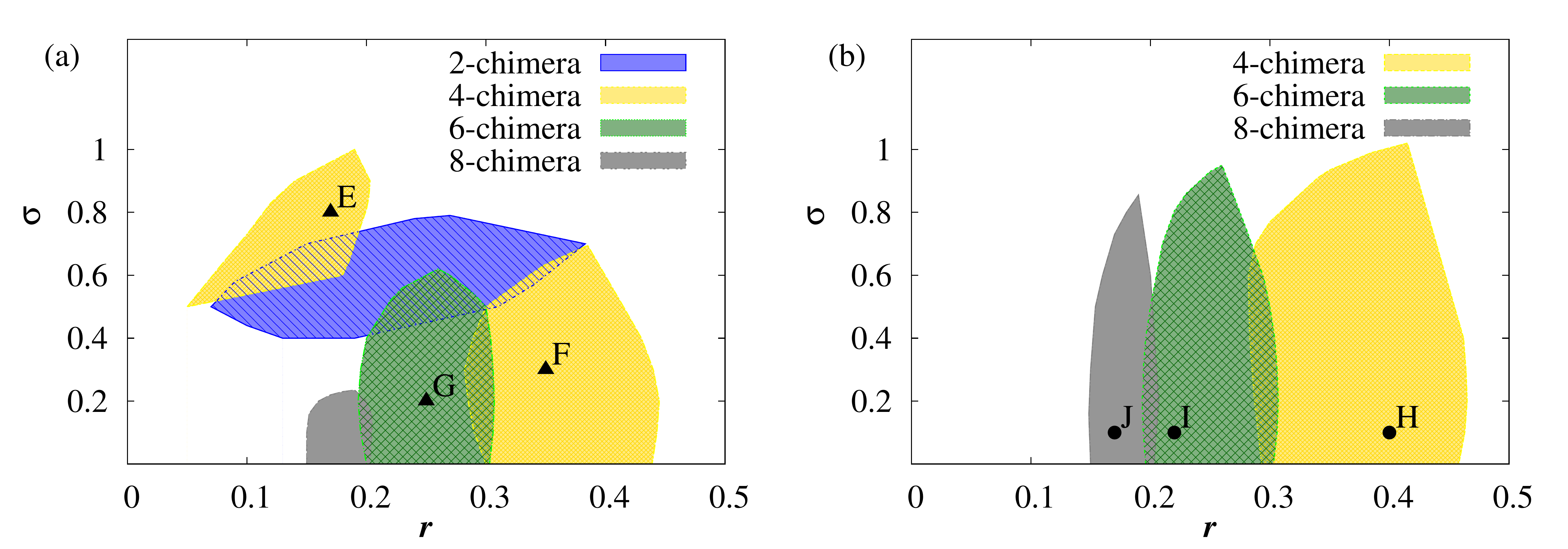}
\caption{Stability regimes for multiple chimera states. (a)~$\varepsilon =0.8$, black triangles marked by
E-G show parameter values corresponding to the panels (a)-(c) in Fig.~\ref{fig4}; (b)~$\varepsilon =1.5$, black circles
marked by H-J denote parameter values corresponding to the panels (a)-(c) in Fig.~\ref{fig5}. Other parameters as in
Fig.~\ref{fig1}.}
\label{fig3}
\end{figure*}
%*******************************************************

Figure~\ref{fig2}(b) depicts the stability regimes for chimera states in the system~(\ref{Eq2}) with~$\varepsilon=0.4$. Compared to the case of ~$\varepsilon=0.2$, the regimes for the chimera states with multiple incoherent domains become larger, and chimera states in this case can be obtained for a wider range of coupling strength~$\sigma$. 
%%%
The reason why the regions in Fig.~\ref{fig2}(b) ($\varepsilon=0.4$) are larger than those in Fig.~\ref{fig2}(a) ($\varepsilon=0.2$) is related to the following qualitative argument: If the coefficient $\varepsilon$ of the nonlinear term in Eq.~(1) is increased, the coefficient $\sigma$ of the coupling term has to be scaled up accordingly to balance the nonlinear term. Hence as $\varepsilon$ is increased the chimera regions extend to larger values of $\sigma$.
%%%

%*******************************************************
\begin{figure*}[ht!]
\includegraphics[height=0.8\linewidth, angle=270]{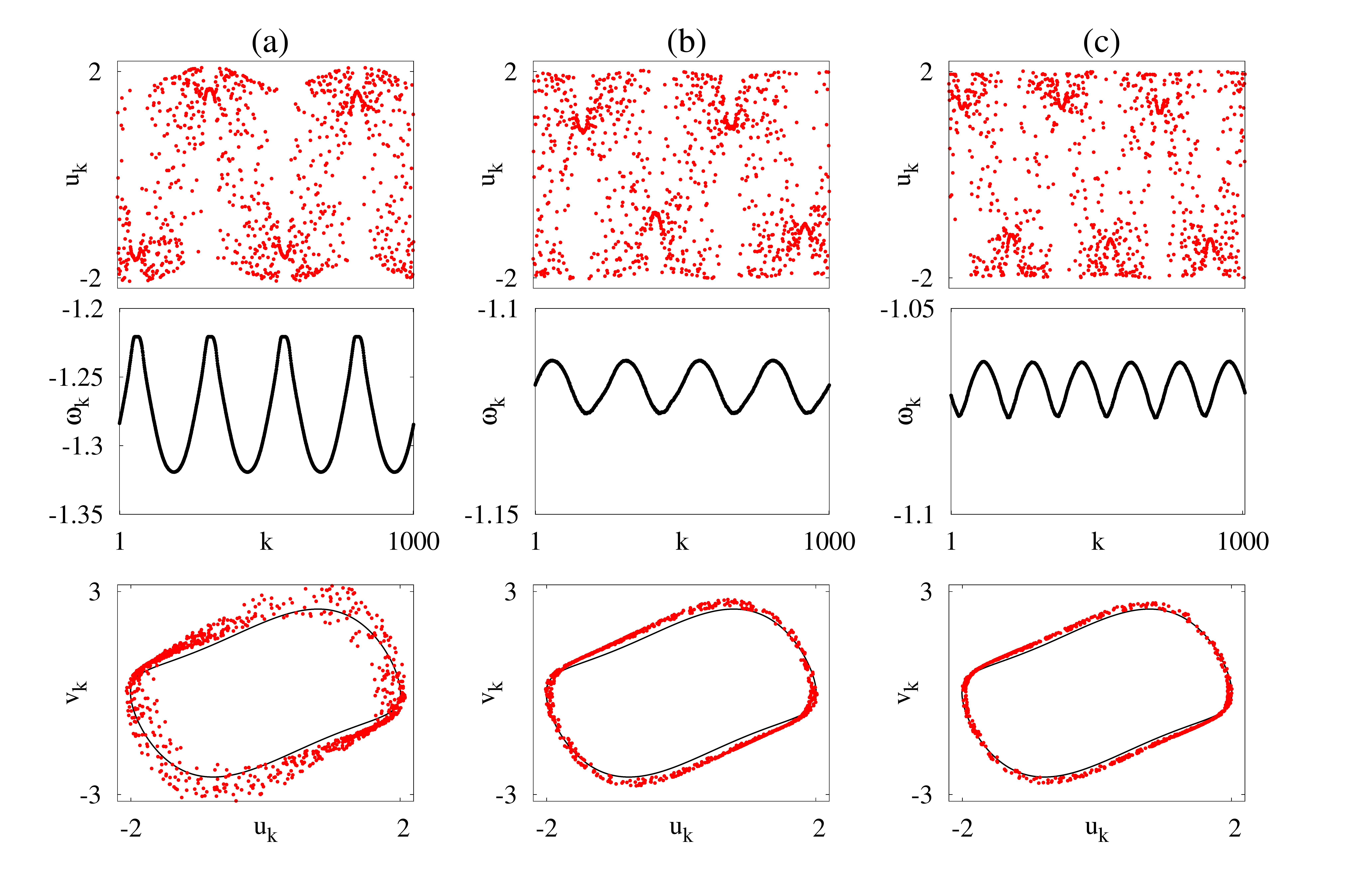}
\caption{Snapshots of the variables $u_k$ (upper panels), mean phase velocities $\omega_k$ (middle
panels), and snapshots in the phase space $(u_k,v_k)$ (bottom panels, limit cycle of the uncoupled unit shown black). 
(a)~$r=0.17$, $\sigma=0.8$, (b)~$r=0.35$, $\sigma=0.3$, (c)~$r=0.25$, $\sigma=0.2$.  Other parameters as in
Fig.~\ref{fig3}(a).}
\label{fig4}
\end{figure*}
%*******************************************************

When the parameter~$\varepsilon$ is increased, the limit cycle of the individual uncoupled Van der Pol oscillators
deforms, and the dynamics on the cycle becomes of slow-fast type. Further increasing $\varepsilon$ leads to strongly
nonlinear relaxation oscillations. This will be discussed in the following.

%*******************************************************
\begin{figure*}[ht!]
\includegraphics[height=0.8\linewidth, angle=270]{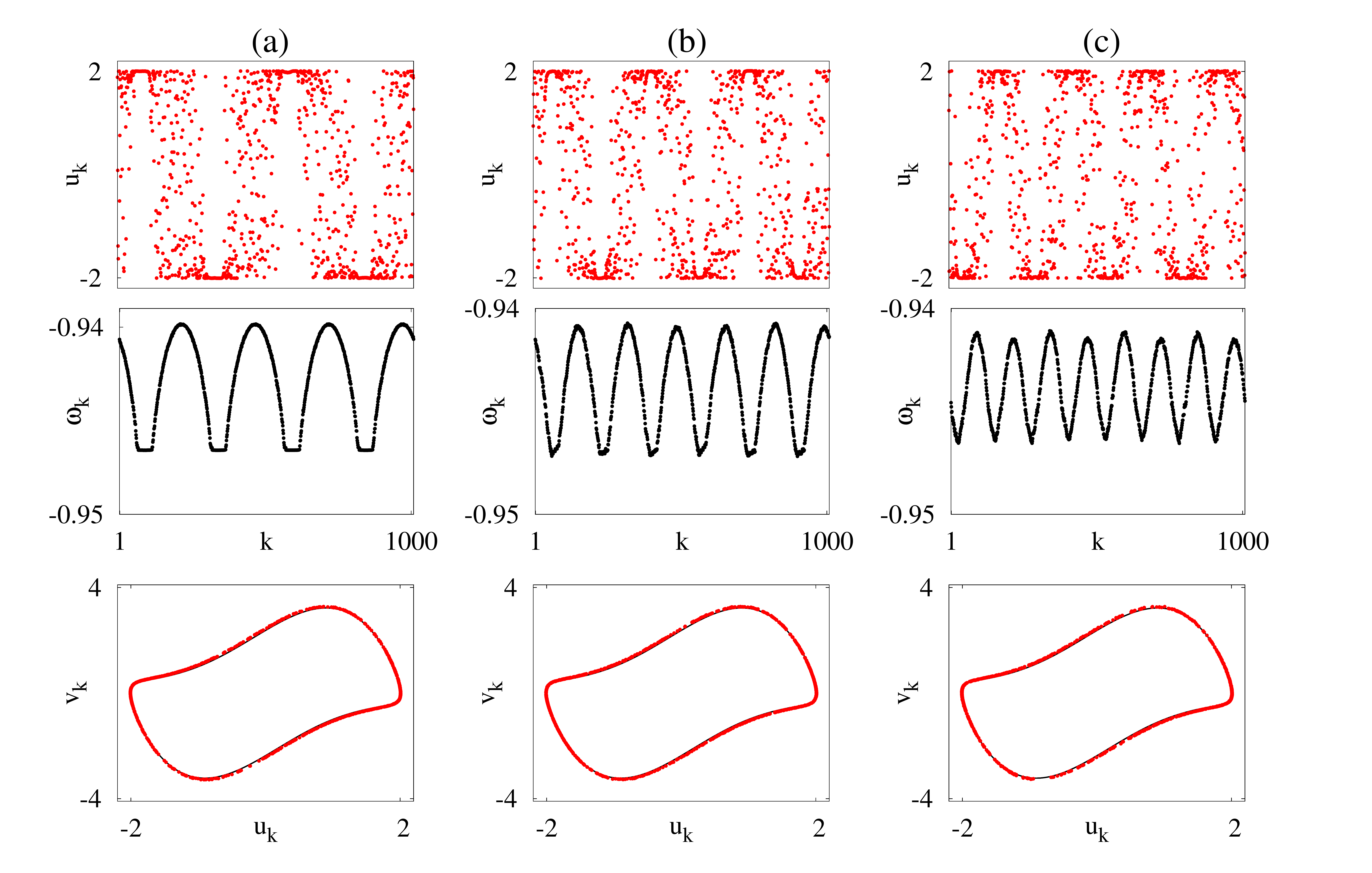}
\caption{Snapshots of the variables $u_k$ (upper panels), mean phase velocities $\omega_k$ (middle
panels), and snapshots in the phase space $(u_k,v_k)$ (bottom panels, limit cycle of the uncoupled unit shown black). 
(a)~$r=0.4$, $\sigma=0.1$, (b)~$r=0.22$, $\sigma=0.1$, (c)~$r=0.17$, $\sigma=0.1$. Other parameters as in
Fig.~\ref{fig3}(b).}
\label{fig5}
\end{figure*}
%*******************************************************

Figure~\ref{fig3}(a) depicts the stability regimes for system~(\ref{Eq2}) with $\varepsilon=0.8$.  Compared to the cases
of smaller $\varepsilon$, there are several qualitative differences in the stability regimes of chimera states. First,
chimera states can be observed for a much larger range of coupling strength~$\sigma$. Second, chimera states with one
incoherent part cannot be observed in the system any more:  for large coupling range we observe chimera states with four
incoherent parts, and furthermore with decreasing coupling range the multiplicity of the incoherent domains of the
chimera states increases.  Black triangles denoted by E, F, and G show values for the parameter pairs $(r,\sigma)$ that
correspond to examples of chimera states depicted in Fig.~\ref{fig4}(a)-(c), respectively.

The peculiarity of the diagram presented in Fig.~\ref{fig3}(a) is the presence of two separate regimes for chimera
states with four incoherent domains. Analyzing this diagram in more detail, one can see that there are two qualitatively
different regions. The first region appears for large coupling strengths, and contains stability regimes for chimera
states with two and four incoherent domains (yellow region containing point E and blue region). These states can be
characterized by strong amplitude dynamics, and the maximum values of the mean phase velocity profile correspond to the
coherent domains of chimera states. The example presented in Fig.~\ref{fig4}(a) [corresponding to point E in
Fig.~\ref{fig3}(a)] depicts these features. Compared with Fig.~{\ref{fig1}} for small~$\varepsilon$, we notice that the
chimera states shown there also show distinct variations along the limit cycle, and the coherent domains of the chimera
states correspond to the maximum in the mean phase velocity profiles.

The second, qualitatively different part of the diagram in Fig.~\ref{fig3}(a) includes three regions for small coupling
strengths (yellow including point F, green, and gray). These regions form a similar sequence with increasing
multiplicity of the chimera starting from four incoherent parts. However, they exhibit  
a qualitative difference. Inspecting Fig.~\ref{fig4}(b) and~(c), which show examples that correspond to the parameter
pairs $(r,\sigma)$ denoted by F and G,
one can notice that the amplitude dynamics becomes weaker in these cases, and the network solution is close to the limit cycle of the uncoupled node shown as black line in the bottom panels. Moreover, the minimum of the mean phase velocities profiles now corresponds to the coherent domains of the chimera states.

%%%
The difference between Fig.~\ref{fig4}(a) on one hand, and Fig.~\ref{fig4}(b) and (c) on the other hand is due to two different types of chimeras.
The 2-, 4-, 6-chimeras in Fig.~\ref{fig2}(a),(b) and Fig.~\ref{fig3}(a)~(point~E), and the chimeras in Fig.~\ref{fig3}(F,G,H,I,J) belong to two different types of chimeras: Fig.~\ref{fig2}(a),(b) and Fig.~\ref{fig3}(a), point E (corresponding to phase portraits shown in Fig.~\ref{fig1}(b),(c),(d) and Fig.~\ref{fig4}(a)) correspond to amplitude-mediated chimeras with strong amplitude-phase coupling, whereas Fig.~\ref{fig3}(a), points F,G and Fig.~\ref{fig3}(b) (corresponding to phase portraits shown in Fig.~\ref{fig4}(b),(c) and Fig.~\ref{fig5}(a),(b),(c)) correspond to pure phase chimeras similar to the ones found for Kuramoto phase oscillators, and since the phase oscillator model can generally be obtained from amplitude-phase models in the weak coupling limit, they occur in the stability diagram (Fig.~\ref{fig3}(a)) only for small coupling strength (points F,G), as opposed to the amplitude-mediated chimeras (point E). This difference is visible in the phase portraits of Figs.~1,4,5 (bottom panels), where the spread of the various oscillators around the cycle of the uncoupled  oscillator (black cycle) is large for amplitude-mediated chimeras, and very small 
for pure phase chimeras where the phase of the cycle is the only dynamical degree of freedom. The difference also shows up in the smaller amplitude variation of the mean phase velocity in the middle panels of Fig.~\ref{fig4}(b),(c) (pure phase chimeras) as compared to Fig.~\ref{fig4}(a), and in the inverted $\omega_k$ profiles: the coherent regions correspond to the minima (Fig.~\ref{fig4}(b),(c)) and maxima (Fig.~\ref{fig4}(a)), respectively.
%%%

Further increase of the parameter $\varepsilon$ of individual Van der Pol oscillators leads to an even stronger
deformation of the limit cycle. In the $(r,\sigma)$ parameter plane the stability regimes for chimera states with four
and more incoherent domains can be observed as shown in Fig.~\ref{fig3}(b) for $\varepsilon=1.5$. The effect of the
coexistence of two qualitatively different types of chimera states is not present there any more, in contrast to the
case of~$\varepsilon=0.8$. Only the second type of the chimera states is observed in the systems now, and the stability
regimes are enlarged towards larger coupling strengths.

Figure~\ref{fig5} depicts examples of multi-chimera states that correspond to the parameter pairs~$(r,\sigma)$ denoted by H,I, and J in Fig.~\ref{fig3}(b). The coherent domains of the chimera states correspond to the minimum of the mean phase velocity profile, and all oscillators stay very close to the limit cycle of the single uncoupled unit, thus the amplitude dynamics of the chimera states in the systems with large~$\varepsilon$ is not as pronounced as in the networks with small~$\varepsilon$.

We conclude that the nonlinearity of the local dynamics indeed strongly influences chimera states in system~(\ref{Eq2}).
The character of the amplitude dynamics, the frequencies of the oscillators belonging to the coherent and incoherent
domains of the chimera states, i.e., the mean phase velocity profiles, and the stability regimes in the coupling
parameter plane undergo a qualitative change with variation of the parameter~$\varepsilon$. Stronger nonlinearity
(larger~$\varepsilon$) results in the dominance of multi-chimera states with weak amplitude dynamics.

\section{Time-delayed coupling}

%*******************************************************
\begin{figure*}[ht!]
\includegraphics[width=\linewidth]{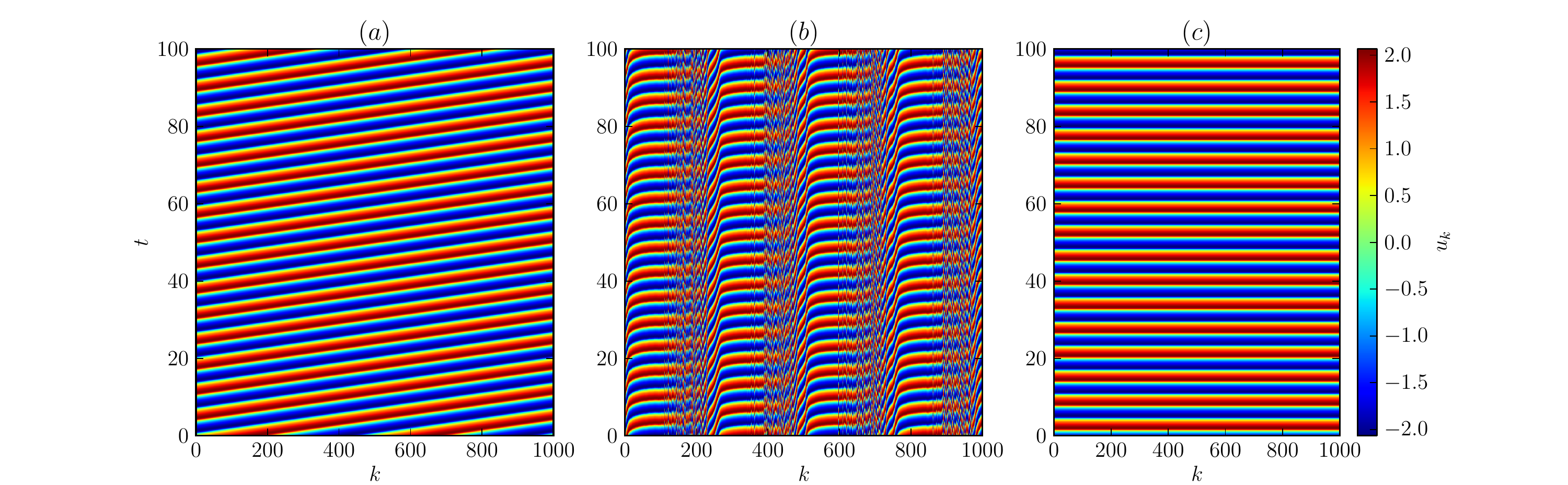}
\caption{Space-time plots of $u_k$ for different values of time delay: (a)~$\tau=1$, (b)~$\tau=3$, (c)~$\tau=6$. Other parameters: $N=1000$, $b_1 = 1$, $b_2 =0.1$, $r=0.4$, $\sigma=0.1$, $\varepsilon =1.5$. Initial conditions as shown in Fig.~\ref{fig5}(a). Transients of 2000 time units are skipped.}
\label{fig6}
\end{figure*}
%*******************************************************

%*******************************************************
\begin{figure*}[ht!]
\includegraphics[height=0.8\linewidth, angle=270]{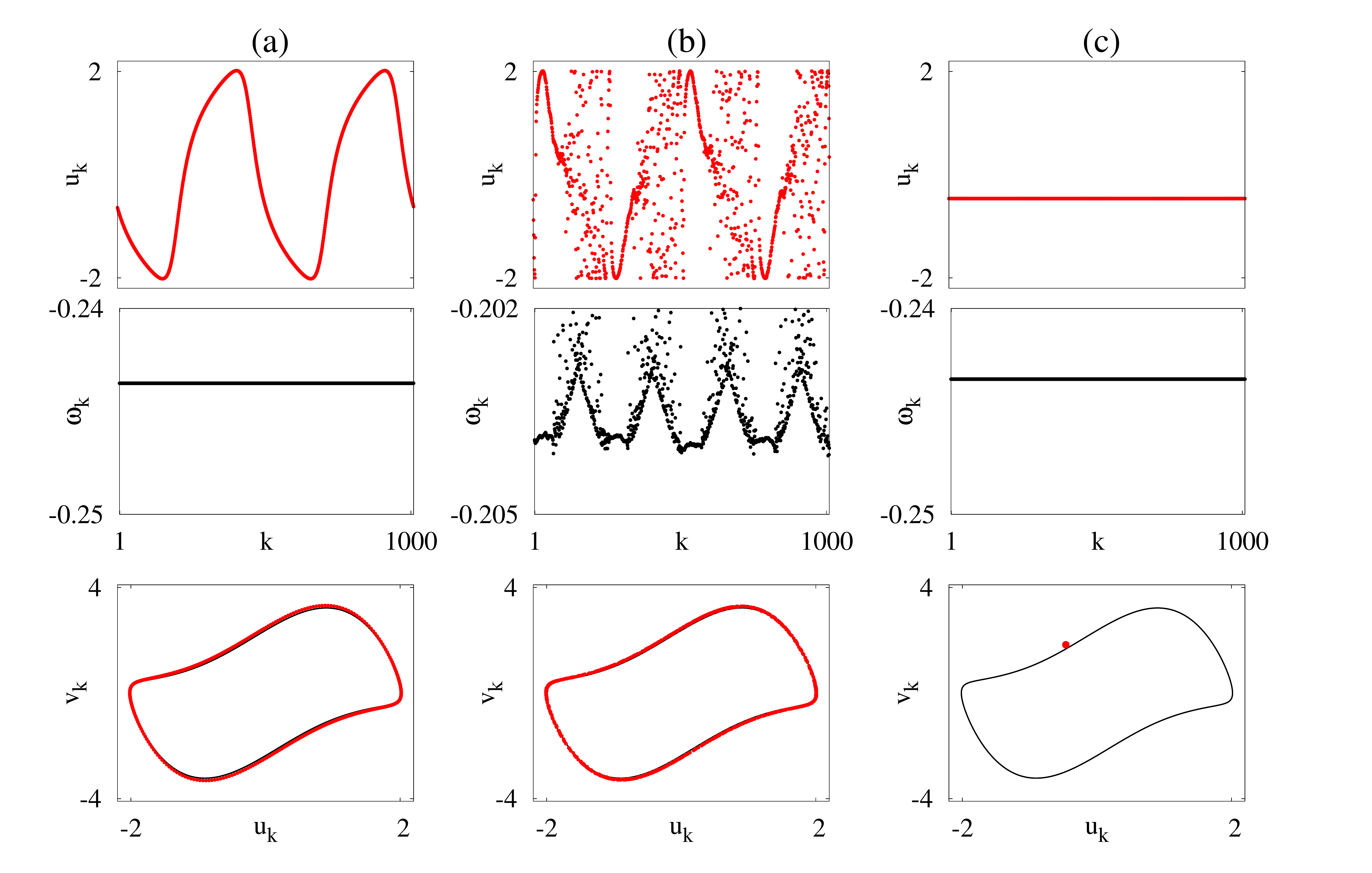}
\caption{Snapshots of the variables $u_k$ (upper panels), mean phase velocities $\omega_k$ (middle
panels), and snapshots in the phase space $(u_k,v_k)$ (bottom panels, limit cycle of the uncoupled unit shown black). 
(a)~$\tau=1$, (b)~$\tau=3$, (c)~$\tau=6$. Other parameters: $N=1000$, $b_1 = 1$, $b_2 =0.1$, $r=0.4$, $\sigma=0.1$, $\varepsilon =1.5$.}
\label{fig7}
\end{figure*}
%*******************************************************
Together with the character of the local dynamics, the coupling between the individual units plays an important role for
the properties of the chimera states. Time-delayed coupling if compared to the instantaneous one represents a more
realistic way to model the interaction between the coupled units. Usually, the coupling range and strength influence the
multiplicity of coherent domains in chimera states. However, it has been shown for phase oscillator networks that time
delay can also induce multi-chimeras \cite{SET08}. The existence of chimera states in systems with time-delayed
couplings has been also reported in~\cite{OME08,NKO13,MA10,SHE09,SHE10}. In particular, for coupled phase oscillator
systems it has been found that chimeras are robust to small time delays and delay distributions \cite{MA10} and can
become unstable depending on the value of delay \cite{SHE09}. Here we consider a model that includes not only phase
but also amplitude dynamics, and show how time delay in the coupling affects chimera 
states that exist in the undelayed system. We demonstrate that by varying the delay value one can both conserve and eliminate chimera patterns.

Let us consider Eq.~(\ref{Eq2}) modified by time-delayed coupling:
\begin{equation}
\begin{aligned}
% \dfrac{d{u}_k}{dt}
\dot{u}_k(t) = & v_k(t)\\
% \dfrac{d{v}_k}{dt}
\dot{v}_k(t) = & \varepsilon [1-u_k^2(t)]v_k(t) - u_k(t) \\
& +\dfrac{\sigma}{2R} \sum\limits_{j=k-R}^{j=k+R} \left\{ b_1\left[u_j(t-\tau)-u_k(t)\right]+ \right.\\
& \left. +  b_2 \left[v_j(t-\tau)-v_k(t)\right]\right\}\\
\end{aligned}
\label{Eq_delay}
\end{equation}
wth $k=1,...,N$ modulo $N$, where $\tau$ is the delay time.

Using the chimera state with four incoherent domains, shown in Fig.~\ref{fig5}(a), as initial condition, i.e., as the
history in the interval $[-\tau,0]$, we fix all system parameters corresponding to this solution, and show exemplary
space-time patterns of Eq.~(\ref{Eq_delay}) for different time delays.  The period of a single uncoupled oscillator is
close to $2\pi$, and we neglect the transients of $2000$ time units. For small time delay ($\tau=1$),  we observe a coherent
traveling wave solution shown in Fig.~\ref{fig6}(a). When the time delay is close to half the oscillation period
($\tau=3$), the chimera pattern is stable and we continue to observe a chimera state with four incoherent domains as
shown in Fig.~\ref{fig6}(b). A larger time delay ($\tau=6$), which is close to the period of a single oscillator, leads
to complete synchronization of all oscillators, see Fig.~\ref{fig6}(c). Figure~\ref{fig7} depicts snapshots of the variable $u_k$ (upper panels), the same snapshots in the phase plane $(u_k,v_k)$ together with the limit cycle of uncoupled oscillator, and the corresponding mean phase velocity profiles (middle panels), for the solutions shown in Fig.~\ref{fig6}.  
%%%
The explanation for the effect of delay is that the delay time interacts with the intrinsic timescale (oscillation period) giving rise to resonance phenomena as found generally for delayed feedback control of steady states, deterministic limit cycles, and noise-induced oscillations, if the delay is an integer multiple or a half-integer multiple of the intrinsic timescale \cite{SCH07}. Delay has a favorable effect on chimeras if $\tau$ is a half-integer multiple, and a favorable (stabilizing) effect on the synchronized oscillations if it is an integer multiple, and may induce traveling waves if it fits with neither condition. In case of the chimera (Fig.~\ref{fig7}(b)), the delay leads to much longer transients, so that with the same length of time interval used for the calculation of the mean phase velocity as without delay, the profiles are more smeared out, but qualitatively similar.
%%%

Our numerical evidence shows similar results for other values of the bifurcation parameter~$\varepsilon$. For chimera states with one or two incoherent domains and sinusoidal character of the oscillations, small delay can lead not only to traveling wave solutions, but also to chimera states with higher number of incoherent domains.

These examples demonstrate that time delay introduced in the coupling can either suppress or preserve the chimera patterns depending upon the value of the delay time relative to the intrinsic oscillation period. 

\section{Conclusion}
In the current study we have demonstrated how the character of the local oscillator dynamics influences chimera states
in networks of nonlocally coupled Van der Pol oscillators. Changing the bifurcation parameter of the single oscillators
allows us to interpolate continuously between sinusoidal and strongly nonlinear relaxation oscillations. We have shown
that nonlinearity facilitates multi-chimera states. 

For small values of the bifurcation parameter~$\varepsilon$ (sinusoidal oscillations) chimera states are characterized by lower multiplicity and more pronounced amplitude dynamics, and the maxima in the mean phase velocity profiles correspond to the coherent domains.  Moving towards the relaxation oscillation regime, with increasing~$\varepsilon$, leads to a 
higher multiplicity of chimera states, but weaker amplitude dynamics. In contrast to the previous case, the coherent domains correspond to the minima of the mean phase velocity profiles, i.e., the profiles are flipped. We have also found that time delay in the coupling strongly affects the chimera patterns in the system, and can lead to chimera suppression and the formation of traveling waves and complete synchronization.

%%%
%The difference between Fig. 4a on one hand, and Fig.4b and c on the other hand is due to two different types of chimeras.
%The 2-, 4-, 6-chimeras in Fig. 2a,b and Fig.3a (E), and the chimeras in Fig.3 (F,G,H,I,J) belong to two different types of chimeras: Fig.2a,b and Fig.3a, point E (corresponding to phase portraits shown in Fig.1b,c,d and fig.4a) correspond to amplitude-mediated phase chimeras with strong amplitude-phase coupling, whereas Fig.3a, points F,G and Fig.3b (corresponding to phase portraits shown in Fig.4b,c and fig.5a,b,c) correspond to pure phase chimeras similar to the ones found for Kuramoto phase oscillators, and since the phase oscillator model can generally be obtained from amplitude-phase models in the weak coupling limit, they occur in the stability diagram (Fig.3a) only for small coupling strength (points F,G), as opposed to the amplitude-mediated chimeras (point E). This difference is visible in the phase portraits of Figs.1,4,5 (bottom panels), where the spread of the various oscillators around the cycle of the uncoupled  oscillator (black cycle) is large for amplitude-mediated chimeras, and very small 
%for pure phase chimeras where the phase of the cycle is the only dynamical degree of freedom. The difference also shows up in the smaller amplitude variation of the mean phase velocity in the middle panels of Fig.4b,c (pure phase chimeras) as compared to Fig.4a, and in the inverted $\omega_k$ profiles: the coherent regions correspond to the minima (Fig.4b,c) and maxima (Fig.4a), respectively.
%%%

We have presented (multi-) chimera states of different type: (i)~pure phase chimeras, which are similar to those found for Kuramoto phase oscillators or weakly coupled amplitude-phase models, and (ii)~amplitude-mediated chimeras with strong amplitude-phase coupling.

Our findings give new insight into the intriguing phenomena of chimera states, and demonstrate that the character of the local dynamics has a strong influence on the chimera patterns in the whole network. These results could be useful from the point of view of applications dealing with different kinds of oscillators, as they can be realized, e.g., in electronic circuits.   

\begin{acknowledgments}
 This work was supported by Deutsche Forschungsgemeinschaft in the framework of Collaborative Research Center SFB 910. 
PH acknowledges support by BMBF (grant no. 01Q1001B) in the framework of BCCN  Berlin. 
\end{acknowledgments}

%\bibliography{ref}
%\bibliographystyle{prsty-fullauthor}

\bibliographystyle{apsrev4-1}

\end{document}